\documentstyle[12pt,epsfig]{article}
\begin{document}

\title{\hfill OKHEP--98--03\\
The Bjorken Sum Rule in the Analytic Approach to \\ 
Perturbative QCD}

\author{K.A. Milton$^{a}$\thanks{E-mail: milton@mail.nhn.ou.edu},
I.L. Solovtsov$^{a,b}$\thanks{E-mail: solovtso@thsun1.jinr.ru},
O.P. Solovtsova$^{a,b}$\thanks{E-mail: olsol@thsun1.jinr.ru} \\ [2mm]
{}$^a${\small Department of Physics and Astronomy,
University of Oklahoma, }\\  
{\small Norman, OK 73019 USA} \\
{}$^b${\small Bogoliubov Laboratory of Theoretical Physics,
 Joint Institute for Nuclear Research,} \\  
{\small 141980 Dubna, Moscow Region, Russia } }

\date{}
\maketitle

\begin{abstract}
Results of applying analytic perturbation theory (APT)
to the Bjorken sum rule are presented. We study the third-order
QCD correction within the analytic approach
and investigate its renormalization scheme dependence.
We demonstrate that, in the framework of the method,
theoretical predictions of the Bjorken sum rule are, practically,
scheme independent for the entire interval of momentum transfer.
\end{abstract}


\newpage

\section{Introduction}

A fundamental test of QCD is obtained by comparing the value
of the quantum chromodynamics (QCD) coupling constant $\alpha_S$
extracted from experimental data at different energy scales
with theoretical predictions given by the renormalization group method.
The corresponding evolution law of $\alpha_S(Q^2)$ is now experimentally
studied down to low momentum transfers, $Q^2 \sim 1\,{\mbox{\rm GeV}}^2$.
The principal experimental information concerning $\alpha_S$
at low $Q^2$ comes from measurements of the inclusive
hadronic $\tau$-decay rate and from deep-inelastic scattering sum rules.
As is well-known, theoretical QCD predictions for these processes
are primarily based on perturbation theory~(PT) improved by the renormalization
group. However, the perturbative expansion is ill-defined at low energies
and the conventional PT method of deriving the
QCD running coupling constant leads to  unphysical singularities
of $\alpha_S(Q^2)$, such as a ghost pole, which are in conflict with the
fundamental principle of causality.
Higher-order PT corrections taken in the asymptotic form
cannot resolve this problem and just add
unphysical cuts. Moreover, at low values of the momentum transfer
there is a strong dependence on the choice of renormalization scheme,
which leads to essential ambiguities in the description of the physical
quantity under consideration.

In~\cite{DVS}, the conventional method of the renormalization
group improvement of the PT approximation has been modified by requiring
K$\ddot {\rm a}$llen-Lehmann analyticity, which reflects the principle of
causality. As a result, the QCD running coupling constant has no unphysical
singularities and at low energy scales behaves significantly differently
than the conventional perturbative one. The method proposed in~\cite{DVS},
called analytic perturbation theory (APT), gives a well-defined procedure of
analytic continuation of the running coupling constant
from the Euclidean (spacelike) into the Minkowskian (timelike)
region~\cite{KS1,KO1}.
This fact allows one to describe the inclusive decay of the $\tau$ lepton
in a self-consistent way~\cite{MSS1}, essential became the correct analytic
properties are important in order to rewrite an initial integral expression
for the $R_{\tau}$-ratio in the form of a contour integral representation.

The renormalization scheme (RS) ambiguity which appears due to
the truncation of the perturbative expansion represents a serious difficulty
in carrying out perturbative calculations. There is no definite solution
to this problem apart from computing indefinitely many terms in the
perturbative expansions.
To somehow avoid this difficulty, various optimization procedures have
been applied, for example, the principle of minimal
sensitivity~\cite{Stevenson81} or the effective charge
approach~\cite{Grunberg84}. The RS dependence can be also reduced
by using the Pad$\acute{\rm e}$ summation method~\cite{J.Ellis96}.
As it has been argued in~\cite{Raczka92,Raczka95}, besides considering
PT predictions in some preferred scheme, one should also study the stability
of those predictions by varying parameters that define the RS over some
acceptable interval. In the framework of the analytic approach the
problem of the RS dependence has been studied for the process of $e^+e^-$
annihilation into hadrons and the process of inclusive $\tau$-decay
in~\cite{SSh97,MSYa}, where it has been demonstrated that the APT approach
can reduce the RS dependence drastically.

Apart from the $\tau$ decay, there is another important observable,
the Bjorken sum rule~\cite{Bj66},
which allows one to extract  $\alpha_S$ at low $Q^2$.
In this paper we analyze the Bjorken sum rule, which is the integral
of the difference between $g_1$ for the proton and neutron,
in the framework of the APT method. We compare our result
with standard PT predictions (see, e.g.~\cite{EK95}) and investigate
the RS dependence of the various theoretical predictions.

\section{The Bjorken sum rule within the analytic approach}

The polarized Bjorken sum rule refers to the integral over all $x$
at fixed $Q^2$ of the difference between polarized structure functions
of the proton $g_1^{\rm p}$ and the neutron $g_1^{\rm n}$,
\begin{equation}
\label{Bj}
\Gamma_1^{{\rm p-n}}(Q^2) \,=\,
\int_0^{1}\,\
\left[g_1^{{\rm p}}(x,Q^2) - g_1^{\rm n}(x,Q^2) \right] dx \, .
\end{equation}
The Bjorken integral~(\ref{Bj}) can be written in terms of
the QCD correction $\Delta_{\rm Bj}$
\begin{equation}
\label{Gamma}
\Gamma^{\rm p-n}_1(Q^2) \,=\,
\frac{1}{6}
\left|\frac{g_A}{g_V} \right| \,
\left[1\,- \, \Delta_{\rm Bj}(Q^2) \right ] \,.
\end{equation}
The value of the nucleon beta decay constant
is taken to be $|g_A/g_V|=1.2601 \pm 0.0025$~\cite{PDG96}.

The perturbative QCD correction to the Bjorken sum rule
in the three-loop approximation
with the use of the $\overline{\rm MS}$ renormalization scheme
and in the massless quark limit has the form
\begin{equation}
\label{Delta_PT}
\Delta_{\rm Bj}^{\rm PT}(Q^2)\,=\,
  \frac{{\alpha}_{\rm PT}(Q^2)}{\pi}
\,+\, d_1  \left[ \frac{{ \alpha}_{\rm PT}(Q^2)}{\pi}\right]^2  \,+ \,
d_2 \left[ \frac{{ \alpha}_{\rm PT}(Q^2)}{\pi} \right]^3  \,,
\end{equation}
where for three active quarks the coefficients are
$d_1^{\overline {\rm MS}}=3.5833$ and $d_2^{\overline {\rm MS}}=20.2153$
\cite{LV91}.
The perturbative running coupling constant
${ \alpha}_{\rm PT}(Q^2)$ is obtained by integration of
the renormalization group equation with the three-loop $\beta$-function.

As it has been demonstrated in~\cite{W78} by using the
Deser-Gilbert-Sudershan representation for the virtual forward
Compton amplitude, the moments of the structure functions are analytic
functions of $Q^2$ in the complex $Q^2$-plane with a cut along the negative
part of the real axis. It is clear that the perturbative
representation (\ref{Delta_PT}) violates these analytic properties due
to the unphysical singularities of the PT running coupling constant
for $Q^2 > 0\,$. To avoid this problem we apply the APT method, which
gives the possibility of combining the renormalization group resummation
with correct analytical properties of the QCD correction to the Bjorken
sum rule.

Let us write down the QCD correction in the form of a spectral representation
\begin{equation}
\label{Delta_ro}
\Delta_{\rm Bj}(Q^2) \, =
\,\frac{1}{\pi}\,
\int_0^\infty\,\frac{d\sigma}{\sigma\,+\,Q^2} \,
\varrho(\sigma) \, ,
\end{equation}
where we have introduced the spectral function, which is defined as
the discontinuity of $\Delta_{\rm Bj}(Q^2)$:
$\,\varrho(\sigma)\,= \,
{\rm Disc}
\left \{\Delta_{\rm Bj}(-\sigma-{\rm i}\epsilon)
\right\}/{2 {\rm i}}\, $.

By calculating the spectral function $\,\varrho(\sigma)$ perturbatively
we get a expression for $\Delta_{\rm Bj}(Q^2)$ which has the correct
analytic properties and therefore has no unphysical singularities.
For instance, the APT running coupling constant
in the one-loop approximation has two terms:
\begin{equation}
\label{APT1}
{ \alpha}_{\rm APT}(Q^2)\,=\,
\, \frac{4\pi}{\beta_0}\,
\left[\, \frac{1}{\ln \left(Q^2/\Lambda^2\right)}\,+\,
\frac{1}{1-Q^2/\Lambda^2} \right ] \>.
\end{equation}
Obviously, the first term in Eq.~(\ref{APT1}) has the standard PT form, but
the second term (which appears automatically from the spectral representation
and restores the correct analytic properties of the running coupling
constant) has an essentially nonperturbative nature. If we rewrite
the second term in Eq.~(\ref{APT1}) in terms of the PT coupling constant,
we obtain an expression which has an essential singularity like
$\exp \left(-{ {4\pi}}/{ {\beta_0 {\alpha}_{\rm PT} } }\right) \, $
as     ${\alpha}_{\rm PT} \to 0\,$.
Therefore, the second term in Eq.~(\ref{APT1}) does not contribute to the
conventional perturbative expansion.
It has been argued in \cite{DVS} that a similar situation holds
also for the running coupling constant in higher order approximations.
The asymptotic PT expression for the running coupling constant is an
expansion in the small parameter $1/\ln(Q^2/\Lambda^2)$.
This approximation violates the $Q^2$-analyticity of the running coupling
constant and does not allow one to describe low energy scales --
the perturbative series diverges in the infrared region.
The APT method removes this difficulty and leads to a quite stable result
for the entire interval of momentum. The difference between the shapes of
the PT and APT running coupling constants becomes significant
at low $Q^2$-scales.

It is convenient to write the three-loop APT approximation
to $\Delta_{\rm Bj}(Q^2)$ as follows
\begin{equation}
\label{Delta_APT}
\Delta_{\rm Bj}^{\rm APT}(Q^2)\,=\,
{\delta_{\rm APT}^{(1)}(Q^2)}
\,+\, d_1 {\delta_{\rm APT}^{(2)}(Q^2)}
  \,+ \,
d_2 {\delta_{\rm APT}^{(3)}(Q^2)}   \,,
\end{equation}
where the coefficients $d_1$ and $d_2$ are the same as
in Eq.~(\ref{Delta_PT}) and
the functions ${\delta_{\rm APT}^{(k)}(Q^2)}$ are derived from the
spectral representation and correspond to
the discontinuity of the $k$-th power of the PT running coupling constant
\begin{equation}
\label{del_APT}
{\delta_{\rm APT}^{(k)}(Q^2)} \,=
\,\frac{1}{\pi^{k+1}}\,
\int_0^\infty\,\frac{d\sigma}{\sigma\,+\,Q^2} \,
 {\rm Im} \;
 \left \{ {\alpha}_{\rm PT}^k(-\sigma -{\rm i}\varepsilon)
 \right\}\,  .
\end{equation}
The function $\delta_{\rm APT}^{(1)}(Q^2)$
defines the APT running coupling constant,
${ \alpha}_{\rm APT}(Q^2) \,=\, \pi \delta_{\rm APT}^{(1)}(Q^2)\,$,
which in the one-loop order is given by Eq.~(\ref{APT1}).

In the case of PT the QCD correction to the Bjorken sum rule is
represented in the form of a power series in ${ \alpha}_{\rm PT}\,$
[see Eq.~(\ref{Delta_PT})], but in the case of APT the same QCD correction
is not a polynomial in the APT running coupling constant.
As follows from Eq.~(\ref{Delta_APT}), the first term of the expansion is
${ \alpha}_{\rm APT}\,/\pi$, but the following terms are not
representable as powers of ${ \alpha}_{\rm APT}\,$.

%
\begin{table}[tbh]
\centering
\caption{ Terms of PT and
APT series for the QCD correction to the Bjorken sum rule.}

\hphantom{}
\label{T1}
\begin{tabular}{ccccccc}  \hline
\hline
\multicolumn{7}{c}
{  $\Gamma^{\rm p-n}_1 (3\,{\mbox{\rm GeV}}^2)\,= \, 0.160  \,$
}   \\ \cline{1-7}
 &   &1{\it st}-term &  & 2{\it nd}-term &  & 3{\it rd}-term
   \\ \hline
$\Delta_{\rm Bj}^{\rm PT}$ & $=$ & $0.131$ & $+$ & $0.062$ & $+$ & $0.045$
   \\ \hline
$\Delta_{\rm Bj}^{\rm APT}$ & $=$ &$0.190$ & $+$ & $0.045$ & $+$ & $0.003$
   \\ \hline   \hline

\end{tabular}
\end{table}

To illustrate the difference between the convergence properties of the
PT expansion (\ref{Delta_PT}) and the APT series (\ref{Delta_APT})
we use, as an example, a typical value of
$\Gamma^{\rm p-n}_1=0.160 \pm 0.014 \,$ at $\,Q^2=3\,{\mbox{\rm GeV}}^2$
taken from~\cite{J.Ellis96}, fixing in such a way the value of the QCD
correction in Eq.~(\ref{Gamma}).
In Table~{\ref{T1} we present numerical results for
contributions to $\Delta_{\rm Bj}$ in different orders for
both the PT and the APT methods.
From this table one can see that the higher order corrections
to the Bjorken sum rule play a different role in the PT and APT approaches.
The convergence of the APT series seems to be much more well behaved
than is that of the PT expansion at such small  $Q\simeq 1.73\,
{\mbox{\rm GeV}}\,$.


\begin{figure}[thp]
\centerline{ \epsfig{file=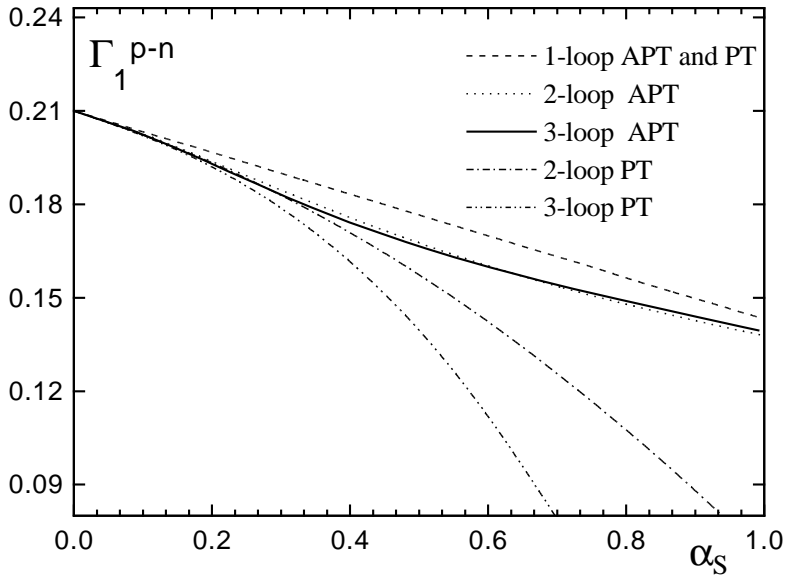,width=10.0cm}}
\caption{{\sl $\Gamma^{\rm p-n}_1\,$  with $1$-, $2$-, and $3$-loop
QCD corrections vs. the coupling constant. }}
\label{bj_fig1}
\end{figure}

Next in Fig.~\ref{bj_fig1}, we show  $\Gamma^{\rm p-n}_1\,$
as a function of the QCD running coupling constant $\alpha_{S}$ in the PT
and APT approaches. As outlined above in the PT case, the function
$\Gamma^{\rm p-n}_1\,$ is an explicit function of the PT running coupling
constant and in the one-loop approximation is represented by a straight
line in Fig.~\ref{bj_fig1}, as a parabola in the two-loop case,
and as a cubic curve in the three-loop one.
At sufficiently large values of $\alpha_{S} \sim 0.5$, the difference
between the 1-, 2-, and 3-loop PT predictions becomes very large.
In the  case of APT, the function $\Gamma^{\rm p-n}_1\,$ only in one-loop
approximation is an explicit function of the APT running coupling constant.
Of course, the coincidence of the one-loop PT and APT curves
in Fig.~\ref{bj_fig1} does not mean that the PT and APT approaches
are physically identical, this is simply a matter of the linear form of
the one-loop approximation -- because the behavior of the PT and APT running
coupling constants are rather different [see Eq.~(\ref{APT1})].
The contribution of the higher loop corrections in the APT case is not
so large as in the PT one and the corresponding curves in
Fig.~\ref{bj_fig1} are quite close to the linear function.
Fig.~\ref{bj_fig1} demonstrates that the APT result is more stable with
respect to higher loop contributions.

%
\begin{table}[tbh]
\centering
\caption{ The QCD parameters extracted from different experimental inputs. }

\hphantom{}
\label{T2}
\begin{tabular}{ccccccc}  \hline \hline
\multicolumn{1}{c}
{ Input [Ref.] }&
\multicolumn{2}{c}
{ Experiment } &
\multicolumn{4}{c}
{ QCD parameters   } \\ \cline{2-7}
 $Q_0^2=10\,{\mbox{\rm GeV}}^2$ & $\Gamma^{\rm p-n}_1(Q_0^2)$  &
 $\Delta_{\rm Bj}(Q_0^2)$  &
 $\Lambda_{\rm PT}$   &
 $\alpha_{\rm PT}(Q_0^2)$ &
 $\Lambda_{\rm APT}$  &
 $\alpha_{\rm APT}(Q_0^2)$  \\ \cline{1-7}
(a)~\cite{SMC_March}
& $0.183 \pm 0.034 $ & $0.129 \mp 0.162$ & $467\,{\mbox{\rm MeV}}$
& $0.275$ & $741\,{\mbox{\rm MeV}}$ & $0.301$
   \\ \hline
(b)~\cite{SMC_Nov}
 & $0.195 \pm 0.029 $ & $0.072 \mp 0.138$ & $138\,{\mbox{\rm MeV}}$
 & $0.177$ & $149\,{\mbox{\rm MeV}}$ & $0.179$
   \\ \hline   \hline
\end{tabular}
\end{table}


\begin{figure}[thp]
\centerline{ \epsfig{file=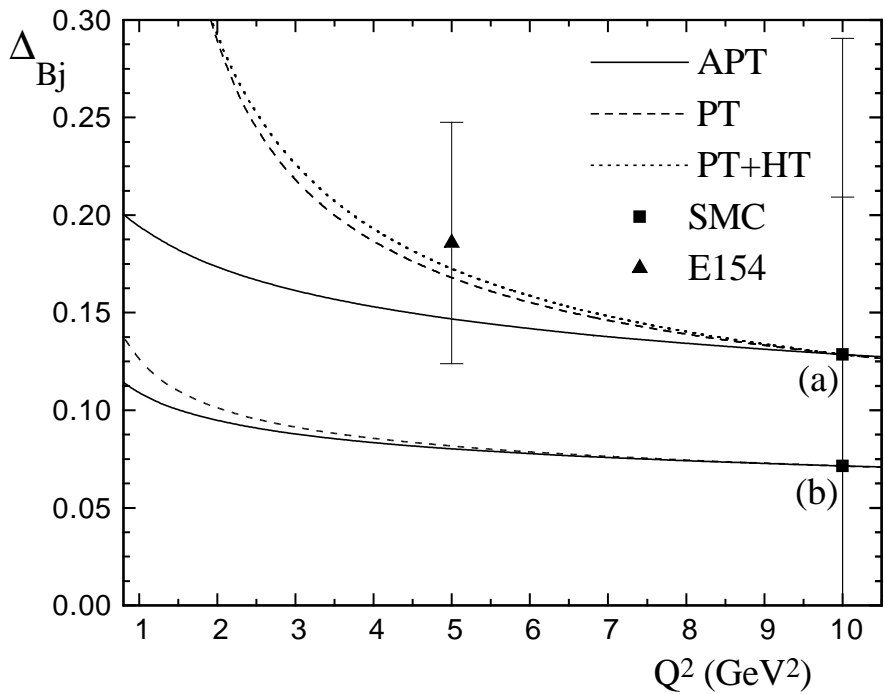,width=11.0cm}}
\caption{\sl The $3$-loop QCD correction to Bjorken sum rule in
the APT and PT approaches at different experimental normalization points. }
\label{bj_fig2}
\end{figure}

To compare the $Q^2$-evolution of the Bjorken sum rule in the APT and PT
approaches, we use as input new experimental values of
$\Gamma^{\rm p-n}_1(Q_0^2=10\,{\rm GeV^2})$ given by the SMC~Collaboration
in Refs.~\cite{SMC_March} and~\cite{SMC_Nov}. These values are presented
in Table~{\ref{T2} as input (a) and (b). The experimental errors of
the present data are too large to determine the QCD correction.
However, to illustrate the evolution, we fix the values of the parameters
$\Lambda$ or the corresponding values of $\alpha_S$ in the PT and APT cases
from the central values obtained from the data  (a) and (b).
By using these normalization points, we plot, in Fig.~\ref{bj_fig2},
the $Q^2$ dependence of the QCD correction $\Delta_{\rm Bj}$.
Besides the normalization points, for illustration, we also represent in
Fig.~\ref{bj_fig2} the recent data of the E154 Collaboration~\cite{E154}
obtained for small $Q^2$. Normalizing at the point (b) we get small
values of $\Lambda$ which are close to each other in the PT and APT cases
(see Table~\ref{T2}); therefore, the corresponding
curves in Fig.~\ref{bj_fig2} are also close to each other.
However, the values of the running coupling constant extracted in such a
way are too small to have  good agreement with other experimental data,
for example, with the data of the E154~Collaboration plotted in
Fig.~\ref{bj_fig2} and with the value of $\alpha_S$ extracted from the
semileptonic $\tau$ decay~\cite{PDG96}.
If the more realistic normalization at the point (a) is used,
the difference between the PT and APT predictions becomes large.
Now, we have the value $\Lambda_{\rm APT}=741\,{\mbox{\rm MeV}}$
[see Table~\ref{T2} input (a)] that is consistent with value extracted
from $\tau$ decay~\cite{MSYa},
$\Lambda_{\rm APT}^{\tau}=871 \pm 155\,{\mbox{\rm MeV}}$.
The discrepancy of the PT value  $\Lambda_{\rm PT}=467\,{\mbox{\rm MeV}}$
and that obtained from the $\tau$ decay,
$\Lambda_{\rm PT}^{\tau}=385 \pm 27\,{\mbox{\rm MeV}}$,
appears to be large.

The dotted curve in Fig.~\ref{bj_fig2} demonstrates that
the inclusion of the higher-twist (HT) corrections does not change
the overall picture of the $Q^2$-dependence of the PT prediction
in the interval under consideration.
However, the higher-twist corrections do reduce the value of $\Lambda$.
By using the higher-twist coefficient
$c_{\rm HT}=-0.03\,{\mbox{\rm GeV}}^2$ (see discussion in~\cite{EK95}),
we obtain the value of
$\Lambda_{\rm PT}=387\,{\mbox{\rm MeV}}$ which agrees well with the
above value of $\Lambda_{\rm PT}^{\tau}$.
Inclusion of such higher-twist corrections is, however, not necessary
in the APT approach to achieve agreement between the QCD scale parameter
$\Lambda$ extracted from the Bjorken sum rule and $\tau$ decay.
Unfortunately, the experimental errors on the Bjorken sum rule are too
large to reach a definite conclusion.

\section{Renormalization scheme dependence}

A truncation of the perturbative expansion leads to some uncertainties
in theoretical predictions for a physical quantity arising from the RS
dependence of the partial sum of the series.
At low momentum scales these uncertainties may become to be very large
(see, for example, an analysis in \cite{Raczka96}).
Thus, it is not enough to obtain a result in some preferred scheme, but
rather it is important to investigate its stability
by varying the parameters that define the RS over some acceptable domain.

Consider the RS dependence of our results.
The coefficients $d_1$ and $d_2$ in Eq.~(\ref{Delta_PT})
are RS dependent. In the three-loop $\beta$-function
\begin{equation}
\label{beta}
\beta(a)=\mu^2\frac{\partial\,a}{\partial\,\mu^2}=-\beta_0\,a^2\,
(1+c_1a+c_2a^2)\, ,
\end{equation}
the coefficient $c_2$ also depends on RS.

In the framework of the conventional approach,
there is no resolution of this problem of the RS dependence apart from
calculating indefinitely many further terms in the PT expansion, and there
is no fundamental principle upon which one can choose one
or another preferable RS.
However, it is possible to define a class of `natural' RS's
by using the so-called cancellation index criterion~\cite{Raczka95}.
According to this criterion a class of `well-behaved' RS's are defined
a such a way that the degree of cancellation between the different terms
in the second RS-invariant~\cite{Stevenson81}
\begin{equation}
\label{rho2}
\rho_2=d_2+c_2-d_1^2-d_1\,c_1
\end{equation}
is not to be very large. The degree of these cancellations can be
measured by the cancellation index~\cite{Raczka95}
\begin{equation}
\label{CI}
C=\frac{1}{|\rho_2|}\left(  |d_2|+|c_2|+d_1^2+|d_1|\,c_1  \right)\, .
\end{equation}

By taking some maximal value of the cancellation index $C_{\rm max}$
one should investigate the stability of predictions for the RS's with
$C\leq C_{\rm max}\,$.
In the case of the $\overline{\rm MS}$-scheme, for the Bjorken sum rule,
the value of the cancellation index is $C_{\overline{\rm MS}}=8 \,$.
We will consider this value as a boundary for the class of `natural' schemes.


\begin{figure}[thp]
\centerline{ \epsfig{file=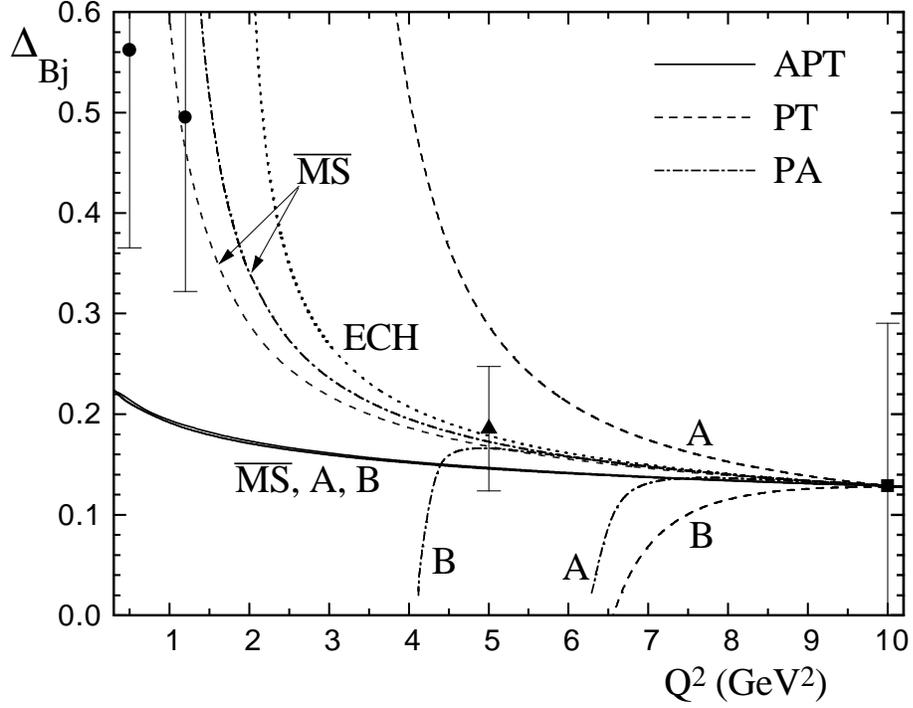,width=11.0cm}}
\caption{\sl Renormalization scheme dependence of predictions for
$\Delta_{\rm Bj}$ vs. $Q^2$ for the APT and PT expansions.
The solid curves, which are very close to each other,
correspond to the APT approach in the $\overline{MS}$, A, B and ECH
schemes. The PT evolution
in $\overline {MS}$, A and B schemes (dash), and in ECH (dot) are shown,
as are the PA results in $\overline {MS}$, A, B schemes (dash-dot).
The SMC data \protect\cite{SMC_March} is denoted by a square, the
triangle is the E154 data \protect\cite{E154}, and
circles are E143 data \protect\cite{E143}.}
\label{bj_fig3}
\end{figure}

Three RS dependent parameters $d_1$, $d_2$ and $c_2$ are
connected by the second
RS-invariant~(\ref{rho2}), and, therefore, any RS is defined by the pair
of numbers $(d_1,c_2)$. In Fig.~\ref{bj_fig3} we plot the PT predictions
represented by dashed lines for three schemes: $\overline{\rm MS}$;
A with parameters $d_1^{\rm A}=-4.3\,$, $c_2^{\rm A}=0\,$,
and with the value of cancellation
index $C_{\rm A}=8\,$; and B with parameters $d_1^{\rm B}=0\,$,
$c_2^{\rm B}=14.5\,$, and $C_{\rm B}=4.3\,$.
As a example of an optimal scheme, we consider the
effective charge (ECH) approach in which $d_1^{\rm ECH}=d_2^{\rm ECH}=0\,$,
$c_2^{\rm ECH}=\rho_2 \,$, and $C_{\rm ECH}=1\,$.
The dotted curve presents the ECH result.
For the sake of illustration we also show the experimental data taken
from~\cite{SMC_March,E154,E143}.\footnote{ If one returns
to Fig.~\ref{bj_fig1} and compares
the properties of convergence of the PT and APT series for the
data~\cite{E143}, where
$\Gamma^{\rm p-n}_1(Q^2\simeq 1\,{\mbox{\rm GeV}}^2)
 \simeq 0.1 \,$,  we can see that the PT series is
not stable with respect to higher loop corrections.}
To normalize all curves,  we used  the experimental value for
$\Delta_{\rm Bj}$ at $Q_0^2=10\, {\mbox{\rm GeV}}^2$  given in
Table~\ref{T2} as input (a).\footnote{To this end, we used $C<8$ for
the B-scheme in order to satisfy the normalization condition. }
The description of experimental data within the
$\overline{\rm MS}$ scheme seems to be quite good; however, as has been
mentioned above, there is no reason why the $\overline{\rm MS}$ scheme
is preferable over, say, the A-scheme. The corresponding series have
the forms $\Delta_{\rm Bj}^{\overline{\rm MS}}  =
x+3.583x^2+20.215 x^3\,$ and
$\Delta_{\rm Bj}^{\rm A} = x-4.36x^2+16.834 x^3 \,,$ where
$ x \equiv \alpha_S(Q^2)/\pi \,$.

Thus the uncertainties coming from the RS dependence of perturbative
calculations are rather large.
At the same time, the APT predictions (solid curves) practically
coincide with each other and, therefore, are RS independent.
This stability reflects the existence of a universal limiting
value~\cite{DVS} of the APT running coupling constant and the small values
of higher loop corrections for the entire interval of momentum.
We evaluated also $\Delta_{\rm Bj}(Q^2)$ using the
Pad$\acute{\rm e}$ approximant (PA)
$[0/2]$ of the PT series (see~\cite{J.Ellis96}).
The results are shown as three dash-dot curves
corresponding to the $\overline{\rm MS}$, A and B schemes.
The PA improves the stability properties, but the sensitivity to the choice
of RS becomes very large for small momentum, $Q^2<5$~GeV$^2$.

Thus the conventional PT prediction at small momentum transfers has a
very large RS ambiguity. The APT approach reduces the RS dependence
drastically. At the three loop level the APT result is, practically,
RS independent.

\section{Summary and conclusion}

We have considered the Bjorken sum rule by using the APT approach
and have demonstrated that the convergence properties of the
APT are much better than are those of the PT expansion.
The APT results have extraordinary stability with respect
to higher loop corrections and also to the choice of the RS.
The analysis performed shows a quite different $Q^2$ evolution
of the Bjorken sum rule in the PT and APT descriptions.
At low $Q^2$ of order a few GeV$^2$, the conventional PT approach
leads to a very rapidly changing function for the Bjorken integral
in many RS's. At the same time, the three-loop APT prediction is
practically RS independent and the $Q^2$ evolution is described
by a slowly changing function.
Unfortunately, the present experimental data for the Bjorken
sum rule have large errors, which does not allow us to
discriminate between the approaches experimentally.
Nevertheless note that experimental data of the E143 Collaboration,
which have just appeared~\cite{E143new}, show, practically, the same value
of $\Gamma_1^{\rm p-n}$ at $Q^2=2,\, 3$ and $5\,{\mbox {\rm GeV}}^2$,
which agrees well with the slow APT $Q^2$ evolution.

The APT approach incorporates ``perturbative'' power corrections
to secure the required analytic properties of the running coupling
constant. These corrections come from the perturbative short distance
analysis and their appearance is not inconsistent with the operator product
expansion~\cite{Grunberg97}. In this note we have concentrated on the
analytically improved perturbative contribution to the Bjorken sum rule
and did not consider higher-twist effects. The higher-twist terms can,
in principle, be potentially important; however, at this stage of the
analysis they lead to an additional uncertainty because the
corresponding values of parameters do not well determined~\cite{Stein}.
Nevertheless, note here that in the framework of the analytic approach
the phenomenological role of ``non-perturbative'' power corrections, which
are controlled by the operator product expansion, is changed, and it will,
in the future, be interesting to consider this problem further.

\section*{Acknowledgement}

The authors would like to thank D.V.~Shirkov for interest in this
work and O.~Nachtmann who brought the paper~\cite{W78} to our attention.
The authors would also like to express their sincere thanks to
L.~Gamberg for helpful discussions.
Partial support of the work by the US National Science Foundation,
grant PHY-9600421,  by the US Department of Energy,
grant DE-FG-03-98ER41066, by the University of Oklahoma,
and by the RFBR, grant 96-02-16126 is gratefully acknowledged.

\end{document}